\documentclass[notitlepage,twocolumn,prl,tightenlines,nofootinbib,superscriptaddress,showpacs]{revtex4}

\usepackage{amsmath}
\usepackage{amssymb,amsfonts}
\usepackage{bm}
\usepackage[mathcal]{euscript}
\usepackage{graphicx}
\usepackage{epsfig}
\usepackage{color}

\begin{document}

\title{Local Coulomb versus Global Failure Criterion for Granular Packings}
\author{Silke Henkes}
\affiliation{Instituut-Lorentz, LION,
Leiden University, P.O. Box 9506, 2300 RA Leiden, Netherlands}
\author{Carolina Brito}
\affiliation{Instituut-Lorentz, LION,
Leiden University, P.O. Box 9506, 2300 RA Leiden, Netherlands}
\affiliation{GIT, SPEC, CEA Saclay, 91191 Gif sur Yvette, France}
\author{Olivier Dauchot}
\affiliation{GIT, SPEC, CEA Saclay, 91191 Gif sur Yvette, France}
\author{Wim van Saarloos}
\affiliation{Instituut-Lorentz, LION, Leiden University, P.O. Box 9506, 2300 RA Leiden, Netherlands}

\begin{abstract}
Contacts at the Coulomb threshold are unstable to tangential perturbations and thus contribute to failure at the microscopic level. How is such a local property related to global failure, beyond the effective picture given by a Mohr-Coulomb type failure criterion? Here, we use a simulated bed of frictional disks slowly tilted under the action of gravity to investigate the link between the avalanche process and a global generalized isostaticity criterion. The avalanche starts when the packing as a whole is still stable according to this criterion, underlining the role of large heterogeneities in the destabilizing process: the clusters of particles with fully mobilized contacts concentrate {\it local} failure. We demonstrate that these clusters, at odds with the pile as a whole, are also {\it globally} marginal with respect to generalized isostaticity. More precisely, we observe how the condition of their stability from a local mechanical proprety progressively builds up to the generalized isostaticity criterion as they grow in size and eventually span the whole system when approaching the avalanche.
\end{abstract}

\maketitle 
Understanding the failure of granular packings is of tremendous importance from both practical and theoretical aspects. Practically, avalanches are clearly of special interest for industrial and natural processes. From a more fundamental point of view, the mechanical rigidity of granular packings is related to the recently explored field of rheology close to dynamical arrest~\cite{Olsson_Teitel,martin10,LiuNagel}, as well as to the nature of the jamming transition for frictional particles~\cite{Deboeuf_EPJB03,Stephanie_PRE05,Lechenault,Makse08}. Despite many studies both from a continuum and a microscopic point of view (see, for example,~\cite{Duran,Mehta00,Daerr,Kabla,Nasuno,Lydie_PRL02,Staron_EPJB05,Lydie_PRE05,Cundall,Nerone}), the mechanisms of failure in frictional granular media are still unclear. 

\begin{figure}[t!]
\centering
 \includegraphics[width = 0.9\columnwidth,trim = 0mm 0mm 0mm 0mm, clip]{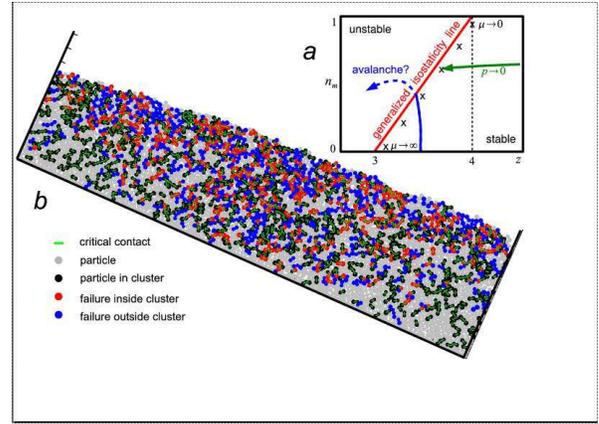}
\caption{
\emph{a} - Generalized isostaticity phase diagram in 2D. In red is the line of marginal stability (Eq.~\ref{eq:geniso}). Frictional packings under isotropic compression unjam -- when compression is released (schematic green arrow) -- at different positions close to this line (crosses) depending on friction $\mu$~\cite{Kostya,DOSslip}. Does a pile slowly inclined under gravity towards avalanche follow the hypothesized blue line? \emph{b} - The system during the avalanche, at an inclination of $24^{\circ}$, for the details of the coloring scheme please refer to Fig.~\ref{fig:clusters_and_corr}.}
\label{fig:schematic_phasediag}
\end{figure}

Macroscopically, the application of the well known Coulomb criterion~\cite{Coulomb} requires the knwowledge of an effective friction coefficient, which remains out of reach of most recent developments. 
From a microscopic perspective, Maxwell derived a global stability criterion based on counting the number of independent contact force components, which has to exceed the number of degrees of freedom for a packing to be mechanically stable~\cite{Tkachenko_Witten}. Recently, this isostaticity criterion has been generalized for frictional packings by including contacts exactly \emph{at} the Coulomb threshold in the above counting argument~\cite{Kostya,Leo_PRE02}. Such fully mobilized contacts are prone to tangential slipping, and it was indeed shown by Staron et al.~\cite{Lydie_PRL02,Lydie_PRE05} that they play a key role in the destabilization process. A frictional packing of $N$ particles with mean contact number $z$ and mean number of fully mobilized contacts per particle $n_{m}$ has only $N d z/2 - N n_{m}$ independent force components, due to the additional restrictions on the tangential forces. For this packing to be stable, this number has to exceed $N d(d+1)/2$, the number of degrees of freedom and the the generalized isostaticity criterion in $d$ dimensions reads:
\begin{equation}
z \geq z_{iso}^{\mu} +\frac{2 n_{m}}{d} \equiv z_{iso}^{gen},
\label{eq:geniso}
\end{equation}
where $z_{iso}^{\mu}=d+1$ is the isostatic value for $\mu=\infty$ frictional packings~\cite{Kostya}. It can be represented in a $(z,n_{m})$ phase diagram (see Figure~\ref{fig:schematic_phasediag}) where a line of marginal stability divides the stable from the unstable regions of phase space. 
Recent molecular dynamics simulations using an isotropic compression protocol~\cite{Kostya} have shown that frictional packings unjam (in the sense that $p\rightarrow 0$) close to the generalized isostaticity line. The final state is characterized by $z(\mu)$ and $n_m(\mu)$ and the linear response properties of these packings suggest that it is the distance to the line of marginal stability, $\delta z^{gen}= z - z_{iso}^{gen}$, which controls stability~\cite{DOSslip}. 

These promising results, regarding both the global stability criterion and the details of the microscopic mechanism, raise two important questions. First, does the stability criterion $\delta z^{gen} > 0$ remain valid in more realistic situations, which necessarily involve finite displacements and anisotropy? More specifically does a granular layer inclined under gravity follow the hypothetical trajectory plotted in blue in the $(z,n_m)$ space on Figure~\ref{fig:schematic_phasediag}? Second, what is the link between the microscopic role of fully mobilized contacts in the failure of the system and the global generalized isostaticity criterion?

In this paper, we explore these two issues using simulation data obtained by Deboeuf et al.~\cite{Stephanie_PRE05}. Having precisely isolated the `avalanches' from the `quiet periods', our first immediate observation is that the pile destabilizes when it is still stable according to the generalized isostaticity criterion. This result prompted us to investigate the microscopic role of the fully mobilized (or critical) contacts. These contacts form elongated clusters, the size of which obeys critical scaling, with a characteristic length which approaches system spanning size near the avalanche onset. These observations, which are consistent with those by Staron et al.~\cite{Lydie_PRL02}, allow us to
investigate the local and global stability properties of these clusters. {\it Local} failure as measured by the number of lost contacts correlates both spatially and temporally with being part of these clusters. Applying then the counting argument at the scale of these clusters, we find that their stability condition builds up progressively from a local mechanical property to the generalized isostaticity criterion as they grow in size. Hence the avalanche can be related to a subset of the packing becoming marginally stable according to the global criterion, while growing up to the system size.
Note however that the above scenario is to be understood as an averaged picture, since these clsuters are  permenantly destabilized and renewed.

The system we study here is a simulation of $2d$ packings of grains under gravity. 
The simulations where performed by Deboeuf et al.~\cite{Stephanie_PRE05} using the contact dynamics code developed by  Staron~\cite{Lydie_these}, which assumes perfectly rigid grains interacting at contacts through a hard core repulsion and a Coulomb friction law: the tangential force at contact, $f_t$, is related to the normal force $f_n$ by the inequality $|f_t| \leq \mu f_n$, where $\mu=0.5$ is the friction coefficient. Beyond the fact that contact dynamics treats them as strictly nonsmooth, these contact laws do not differ from those more commonly used in discrete simulations~\cite{Moreau}.
The system consists of $4000$ circular grains with diameter uniformly distributed between  $[d_{min}, d_{max}]$ in a way to ensure $20\%$ polydispersity. The length of the box is about $120d$ and the height is about $35d$, where $d$ is the mean diameter of the grains. Initially the grains free fall in the box set horizontally. The box is then tilted quasistatically to the desired inclination $\theta$.
Deboeuf et al.~\cite{Stephanie_PRE05} have run several histories of inclination -- including several back and forth oscillations -- to investigate the stress anisotropy.
In the present case, we have selected the final part of the pile history, starting with a horizontal pile with $\theta_0 = 0\,^{\circ}$ and tilting it in the direction of positive $\theta$ until $\theta \approx 30\,^{\circ}$.
We use $50$ independent runs, with different initial conditions and our results are usually shown as an average over these runs. For each run, particle positions, contacts and forces were stored in successive frames separated by $50$ computational time steps, which corresponds to an angle variation of $\delta \theta = 0.05\,^{\circ}$.

\begin{figure}[t!]
\centering
\includegraphics[trim = 5mm 0mm 10mm 5mm, clip, width=0.48\columnwidth]{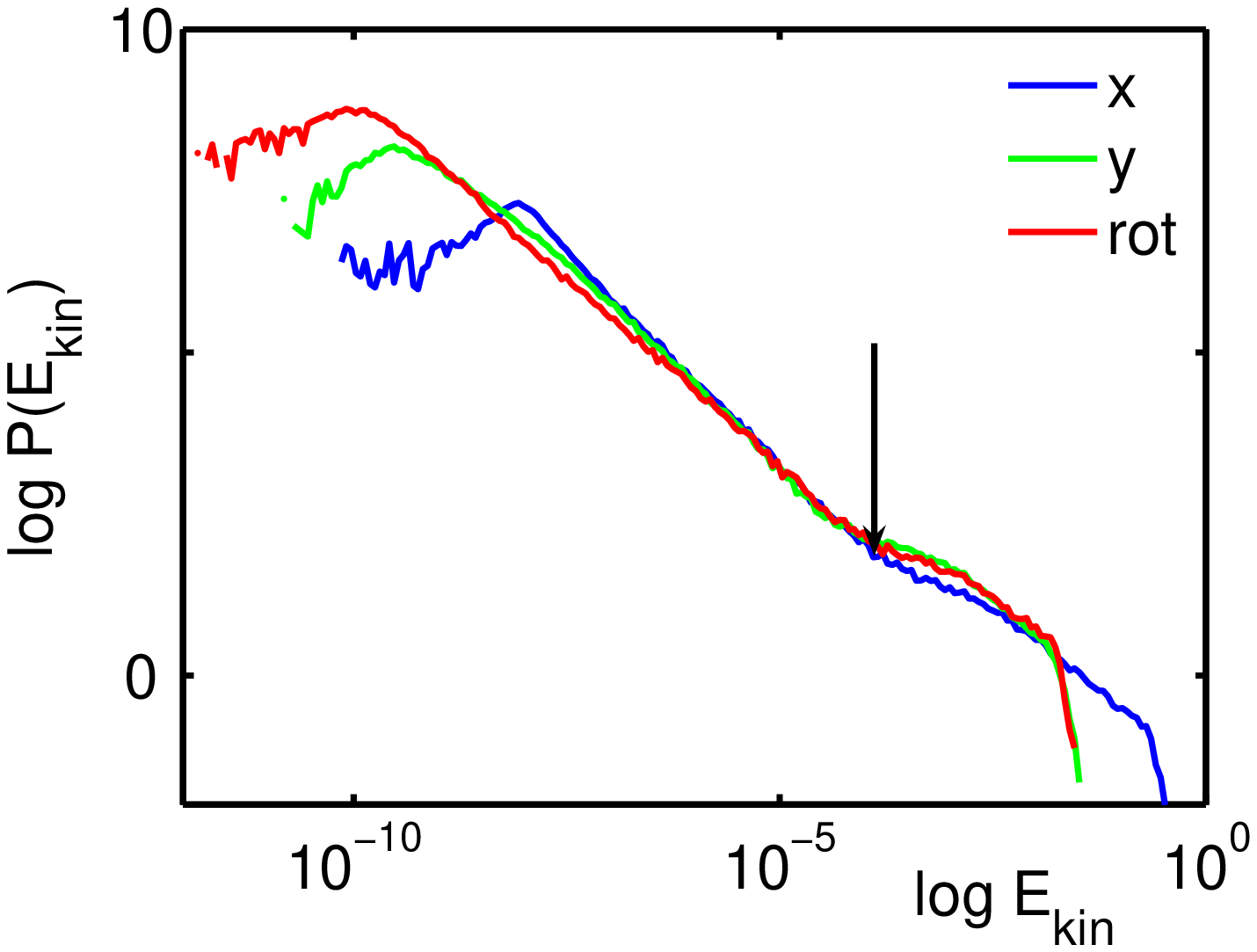}
\includegraphics[trim = 5mm 0mm 10mm 5mm, clip, width=0.48\columnwidth]{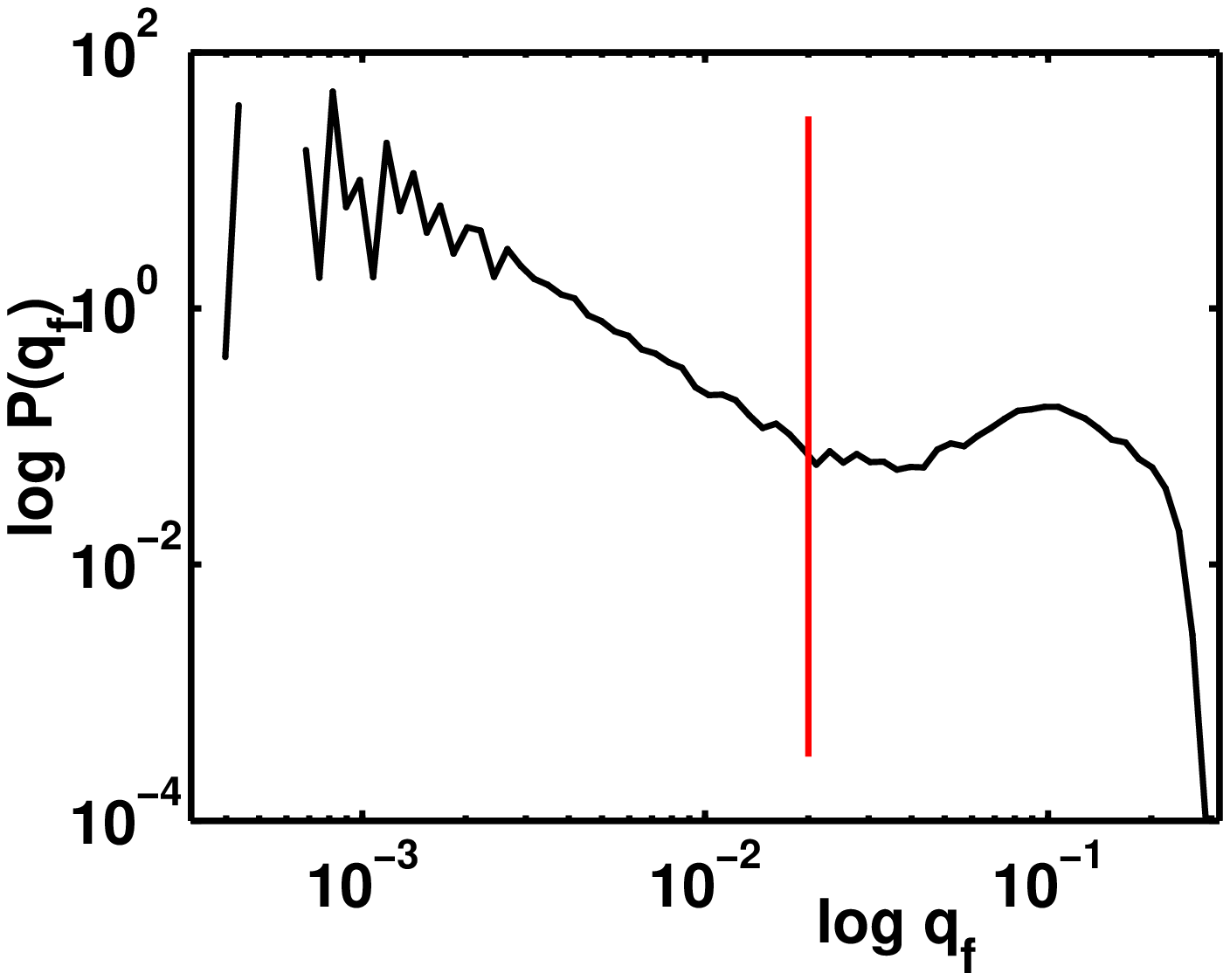}
\includegraphics[trim = 5mm 0mm 10mm 5mm, clip, width=0.48\columnwidth]{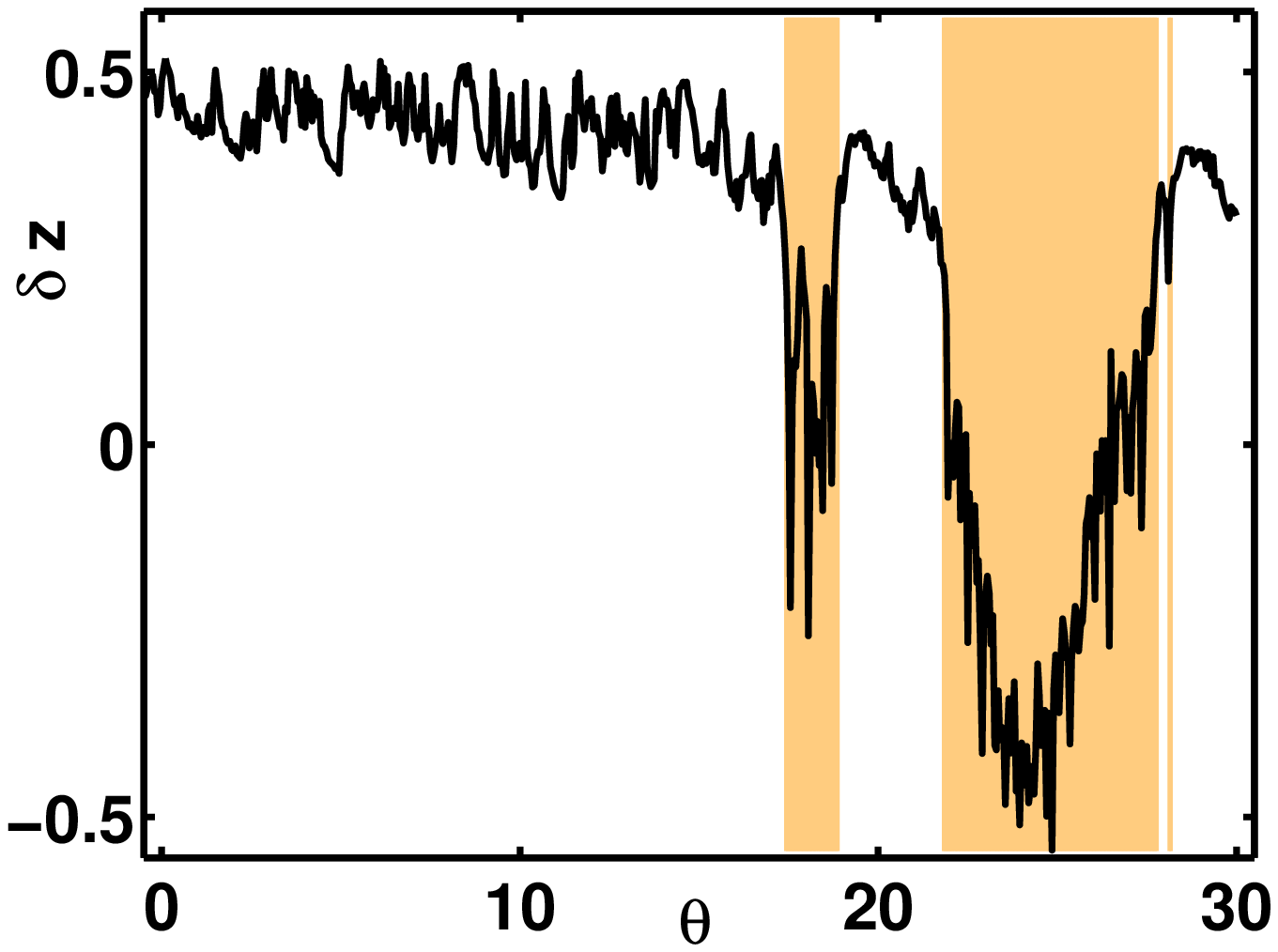}
\includegraphics[trim = 5mm 0mm 10mm 5mm, clip, width=0.48\columnwidth]{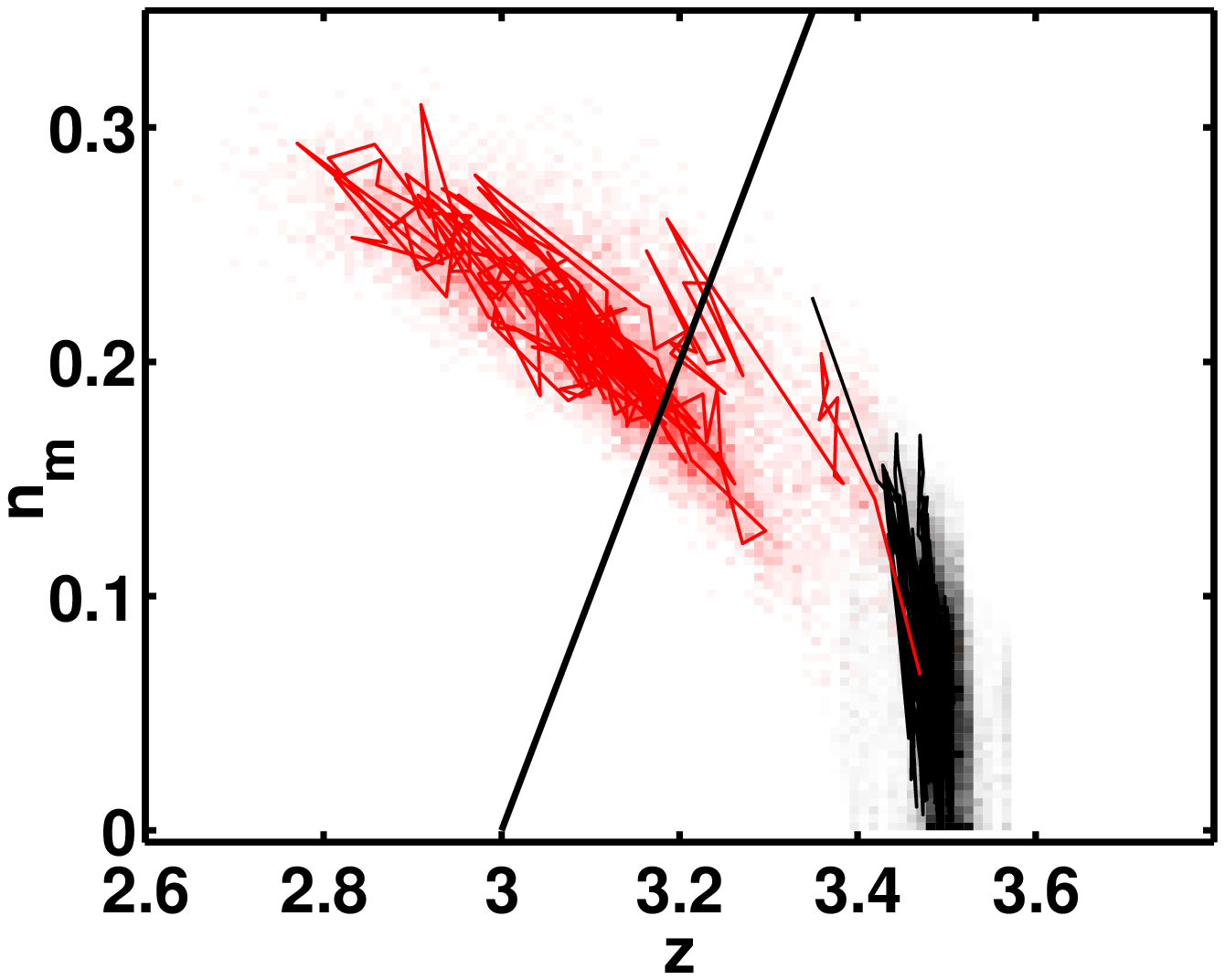}
\caption{\emph{a} - Distribution of the kinetic energy of the system averaged over all frames and all runs. \emph{b} - Avalanche criterion: Probability density of the fraction of contacts lost between frames (log-binning). The criterion at $2\%$ of failing contacts is shown in red. \emph{c} - Distance from generalized isostaticity for a sample run; the run is divided into quiet and avalanching periods (in orange). \emph{d} - Probability density of the position of the system in the generalized isostaticity diagram, overlaid by the trace of a sample run. The quiet period is shown in black, while the avalanching period is in red. The system clearly loses stability before crossing the marginal stability line, and the dynamics during the quiet period is dominated by the fully mobilized contacts.}
\label{fig:av_criterion}
\end{figure}

Let us first turn to the identification of the avalanche periods. We are interested in the properties of the packing just before failure occurs, i.e. for $\theta\rightarrow\theta_{m}$, where $\theta_{m}$, the failure angle is, known to show considerable statistical variance~\cite{Stephanie_PRE05}. 
Obtaining a good criterion to precisely identify the start of an avalanche is not straightforward: one possibility with limited resolution is to consider the probability distribution of the kinetic energy (either translational and rotational) and identify as a threshold the value above which the distribution escapes the power law reported in the quiet region (see the arrow in Figure~\ref{fig:av_criterion}a). A more explicit signature of failure is given by considering the probability density of the fraction of contacts lost between two subsequent snapshots. As observed in Figure~\ref{fig:av_criterion}(b), the data clearly fall into two distinct sub-populations (note the logarithmic scale on the x-axis), which we associate with the quiet period (left) and the avalanching period (right). The good separation between the two populations allows us to apply a cutoff of a maximum of $2\%$ of contacts lost between frames to determine the end of the quiet period (in red). In Figure~\ref{fig:av_criterion}(c), this criterion has been applied to a sample configuration, and it is clearly able to capture the location of significant events in the packing. We have checked the robustness of this criterion by varying the cutoff between $1\%$ and $4\%$, which typically changes the location of the onset of the avalanche by 1-2 frames, corresponding to an error in $\theta_{m}$ of $0.05^{\circ}-0.1^{\circ}$. We have observed that the position of the avalanche onset is very variable, and that there may be several avalanches of various sizes within a single run. In the following all ensemble averages are performed as a function of $\theta_{m}-\theta$, where $\theta_{m}$ is the location of the next avalanche onset. 

Figure~\ref{fig:av_criterion}d shows the probability density of the system in the parameter space $(z,n_m)$, together with the trace of one sample run. The black (respectively the red) symbols correspond to the quiet (resp. avalanche) period. Two features are apparent. First, the quiet period and the avalanche occupy very well separated regions of the parameter space and the transition between them is abrupt. During the quiet period, the mean contact number $z$ is quenched, and all of the dynamics is due to fully mobilized contacts appearing and disappearing in the packing (corresponding to an up-and-down motion in the graph). This is confirmed by our observation that throughout the pre-avalanche period, particles move typically less than $1\%$ of their diameter and remain ``caged''. When the system enters the avalanche period, the structure of the packing breaks down and the trace moves along a diagonal. 
Second, the system crosses the marginal stability line only \emph{after} the start of the avalanche. 
This means that the generalized isostaticity criterion, (Eq.~\ref{eq:geniso}), is a necessary criterion for stability, but not a sufficient one.

\begin{figure}[t!]
 \centering
\includegraphics[width=0.85\columnwidth,trim = 0mm 0mm 5mm 5mm, clip]{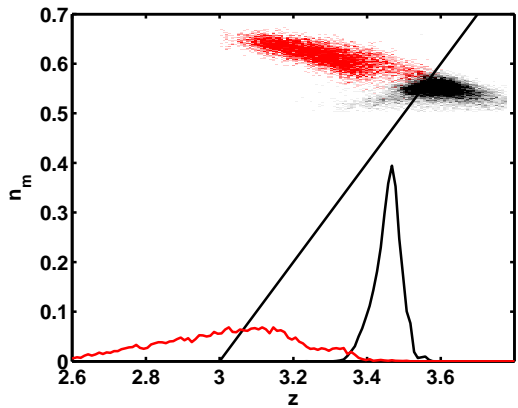}
\caption{Probability density of the position of the clusters in the generalized isosaticity diagram, before the avalanche in black, and during the avalanche in red. The clusters straddle the marginal stability line during the quiet period. The z-distribution of the remainder of the system is shown below - note that since $n_{m}=0$ for this subset, we have $z_{iso}=3$.}
\label{fig:geniso_diag}
\end{figure}

Generalized isostaticity is a global, mean-field type criterion, and it is likely to overestimate the stability of the packing if the system behaves in an heterogeneous way. We have checked that such heterogeneities are not related to the geometry. If we exclude the top and bottom single layers of particles, which cannot easily be integrated into the counting, no part of the system is critical at the start of the avalanche (note that we only used the bulk of the system for Figure~\ref{fig:av_criterion}d). We also have cut the system both into lanes, and several other patterns and found that the same general conclusion holds: within a scale of a few frames, the avalanche onset measured by this method is the same for all sub-parts of the system ensuring that the avalanche is truly a global property of the system (see also Figure~\ref{fig:geniso_diag}b). Note that for a granular bed much deeper than our relatively shallow layer of particles, one may observe a stronger dependance of $n_{m}$ on depth than in the present case and that the system could then eventually separate.

\begin{figure}[t!]
\centering
\includegraphics[trim = 5mm 30mm 10mm 2mm, clip, width=0.9\columnwidth]{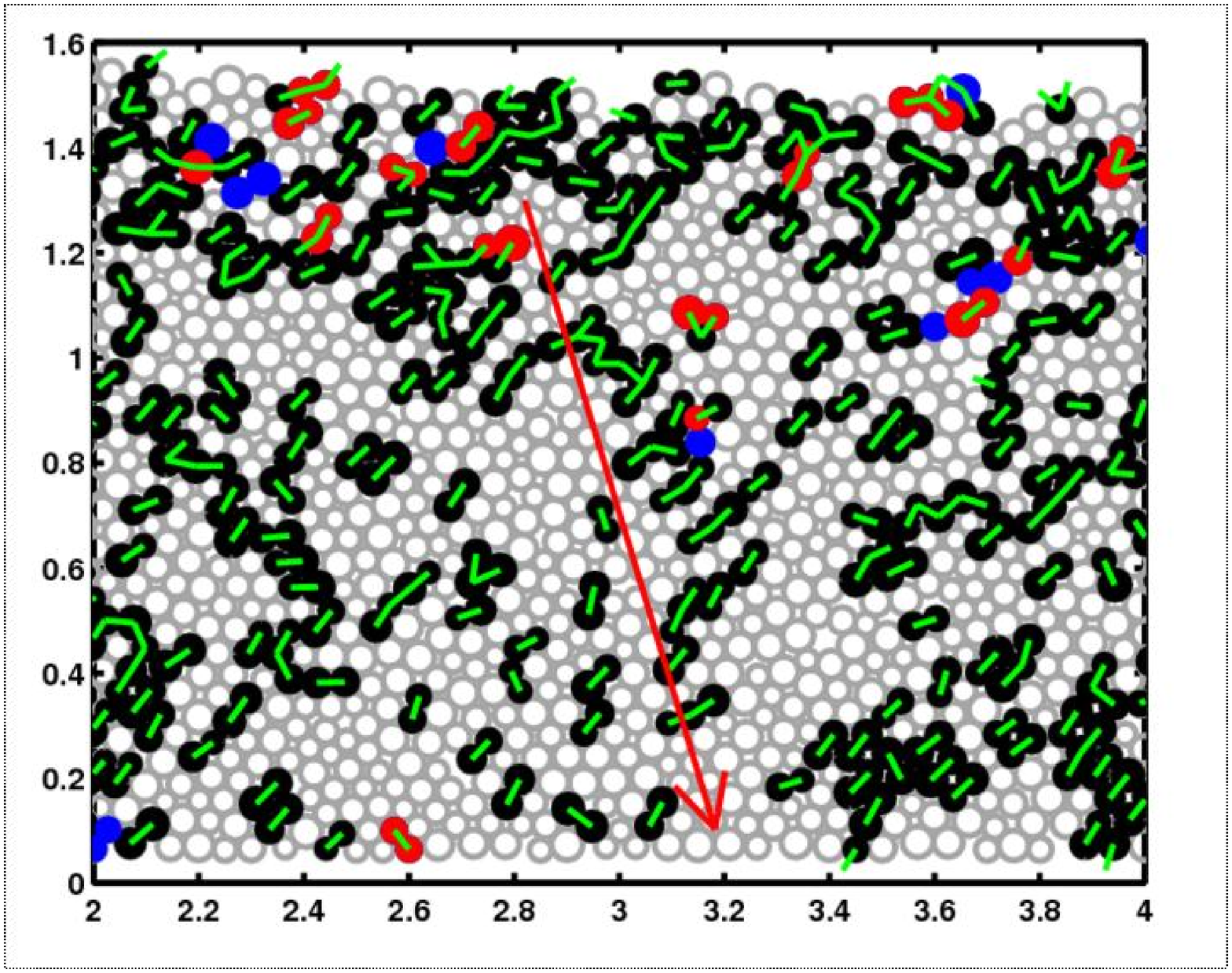}
\includegraphics[trim = 5mm 2mm 7mm 5mm, clip, width=0.45\columnwidth]{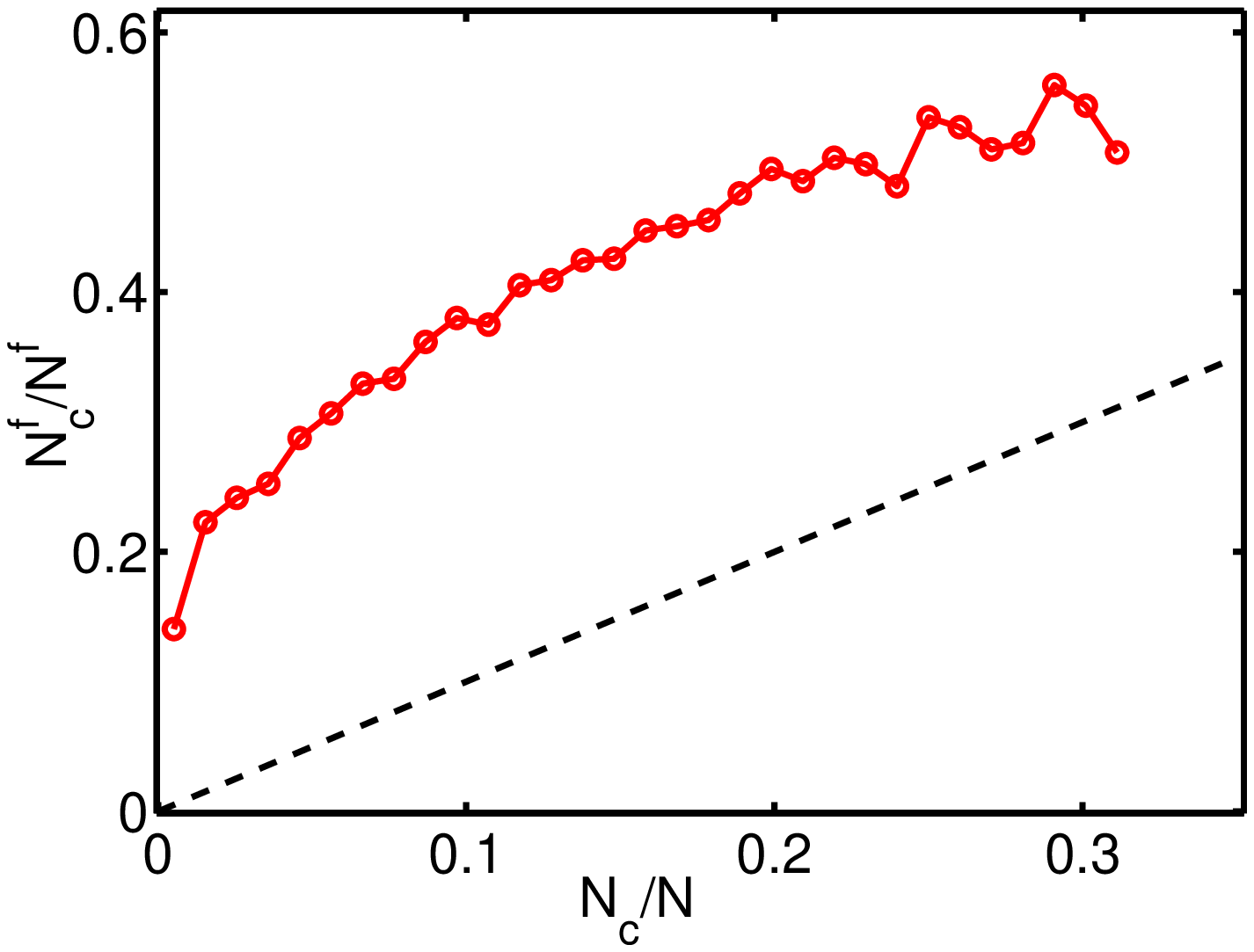}
\includegraphics[trim = 5mm 0mm 5mm 5mm, clip, width=0.45\columnwidth]{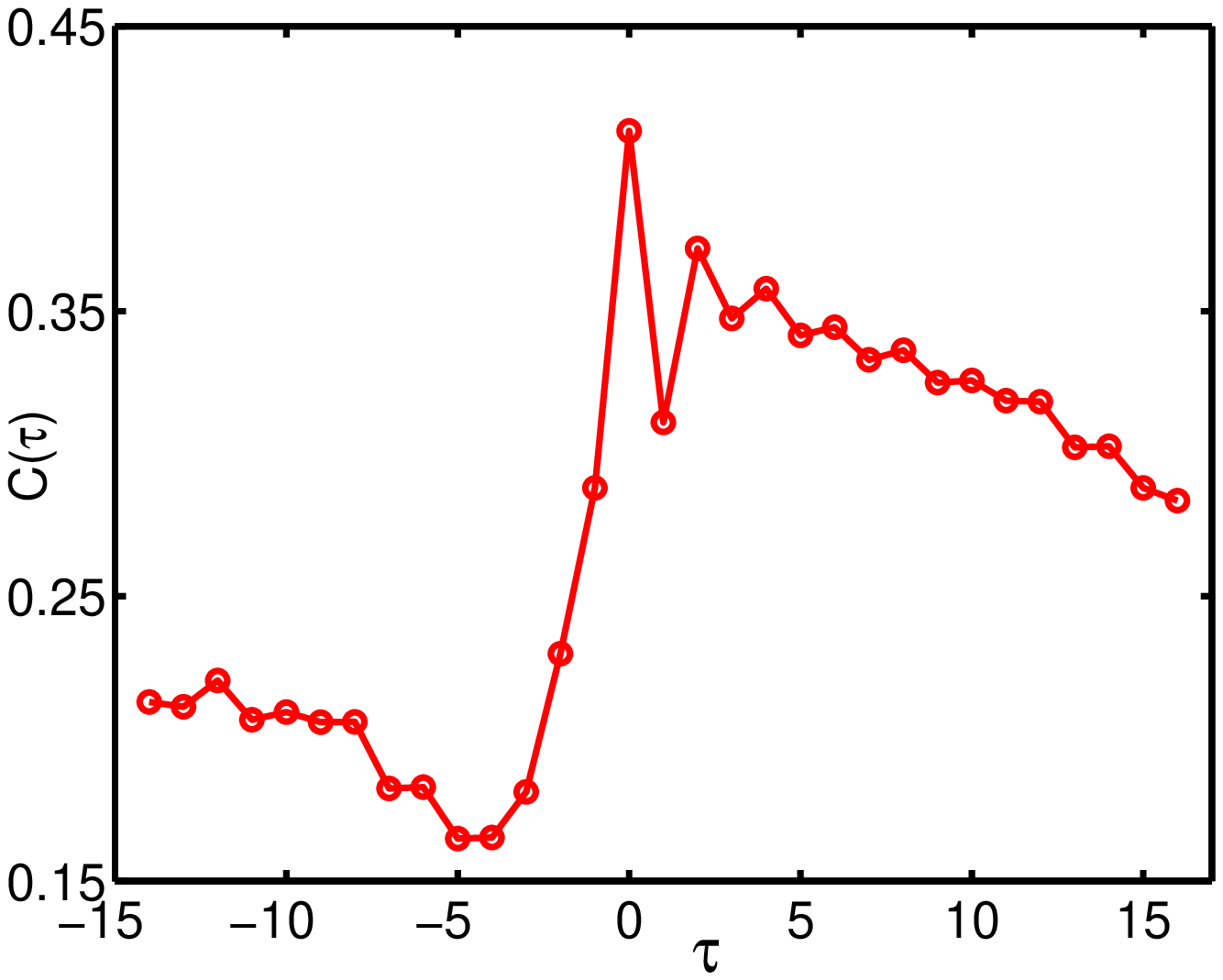}
\caption{(Color online) Top: Contact properties close to the avalanche onset, $\theta_{m}-\theta\approx 0.3^{\circ}$; clusters of particles with at least one critical contact (in green) are in black. Particles with failing contacts belonging to a cluster are red, failure of the other particles is in blue. The red arrow indcates the direction of gravity -- note the directionality of the clusters. Bottom: Spatial (left) and temporal (right) correlation between clusters and failure during the quiet period. Left: Fraction of failure which occurs in a cluster, $N^{f}_{c}/ N_{c}$ as a function of fraction of the system in a cluster, $N_{c}/N$. The dashed line is the limit where failure and clusters are uncorrelated. 
Right: Temporal correlation between the maximum cluster size at time $t$ and the number of particles that fail between $t$ and $t+\tau$; the units are frames each corresponding to $\Delta \theta = 0.1^{\circ}$.} 
\label{fig:clusters_and_corr}
\end{figure}

Inspired by Staron's work~\cite{Lydie_PRL02}, we identify the critical contacts as a natural candidate for the source of the heterogeneity suggested by the above observations. Since the contact number is quenched during the quiet period and all changes in $\delta z^{gen}$ are borne by the critical contacts, while the remainder of the system does not contribute, we can divide the system into two subpopulations: those particles with at least one fully mobilized contact, and those without.
Figure~\ref{fig:clusters_and_corr}, top, illustrates the critical contacts close to the onset of the avalanche. It becomes apparent that the particles with critical contacts organize in rapidly fluctuating \emph{clusters}. These clusters are rather anisotropic and their preferential direction of alignment makes an angle of roughly $60^{\circ}$ with respect to gravity. This is in agreement to the Mohr-Coulomb failure criterion and its extension in the context of elasto-plasticity~\cite{Nederman}, which would predict a failure angle of $1/2 \left(\pi/2-\text{atan}(\mu)\right) \simeq 30^{\circ}$, that is along the slip plane at the critical contacts, normal to the contact orientation.

We now turn to the relation between these clusters and {\it local} failure, as measured by the contacts lost between two consecutive frames.  We find both spatial and temporal correlations between clusters and failing contacts. As illustrated in Figure~\ref{fig:clusters_and_corr}, particles within clusters are more likely to have failing contacts. The particles with failing contacts which belong (don't belong) to a cluster are illustrated in red (blue) in the top figure, and even though less than half the system is part of a cluster, most of the failure occurs in clusters. The bottom-left figure confirms that the fraction of particles which fail \emph{and} belonged to a cluster in the previous frame as a function of the fraction of the system which is part of a cluster, is much higher than if both events were uncorrelated (dashed line). Similarly, we track the size $S_{m}$ of the largest cluster in the system and correlate it with the number of contacts $N^{f}$ that fail in the following frames using
\begin{equation} 
C(\tau) = \frac{ \sum \left (S_{m}(k)-\langle S_{m}\rangle \right) \left( N^{f}(k\!+\!\tau) - \langle N^{f} \rangle \right )}{({Var(S_{m}) Var(N^{f}))^{1/2}}}, 
\label{eq:corrdef}
\end{equation}
where the limits on the sum are chosen such that both $k$ and $k\!+\!\tau$ fall within the quiet period. The mean and the variance are computed over the same set of frames. As shown in Figure~\ref{fig:clusters_and_corr} bottom-right, both quantities are correlated during the whole quiet period, but the maximum correlation happens at $\tau=0$: just after the cluster size peaks, the failure rate is enhanced by about $40\%$. The slow decay for $\tau>0$ is due to the persistence of clusters, since most of critical contacts tend to belong to particle pairs which repeatedly form and loose contact. 
In contrast, during the avalanche, both the correlations between clusters and failure and the directionality of the clusters are lost (see Figure~\ref{fig:schematic_phasediag}) - a signature of the breakdown of the structure of the system.

The clusters of particles with at least one critical contact hence appear as natural candidates for being the seeds of the destabilization process. However, for stability reasons, this is far from trivial. Let us recall here that for one particle to be stable, it must satisfy the local stability criterion $z-2 n_m/d \geq 1/2(d+1)$, which states that the $d z -2 n_m$ force components need to be able to constrain the $1/2 d(d+1)$ degrees of freedom of the particle. The particles which belong to these clusters have by construction a minimal $n_m=0.5$ (at least one critical contact shared by two particles). The average number of contacts for the whole pile is $\langle z\rangle \simeq 3.5$. Thus these particles easily satisfy the local stability criterion. However when such individually stable particles aggregate into clusters, one expects that above some mesoscopic size, the cluster will need to satisfy the global criterion~(\ref{eq:geniso}). Assuming for the moment that the average contact number within the clusters is also close to $\langle z\rangle \simeq 3.5$, one obtains that these clusters are prone to be marginally stable as confirmed in Figure~\ref{fig:geniso_diag}, where one can see that the clusters straddle the marginal stability line during the quiet period (in black) and hence are highly unstable to perturbations. And indeed one observes visually that they are very intermittent, permenantly loosing and gaining particles, vanishing and reforming in contrast with the remainder of the system is always far removed from isostaticity during the quiet period. Hence, it is not clear that the clusters can grow up to a system spanning size.

\begin{figure}[t!]
\centering
\includegraphics[angle=-90, trim = 0mm 0mm 0mm 0mm, clip, width=0.85\columnwidth]{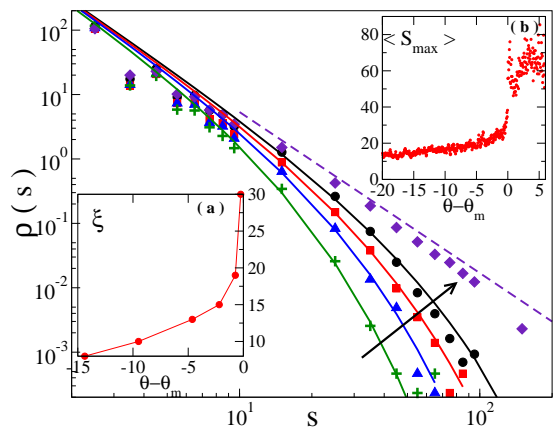}
\caption{(Color online): Cluster size distribution $\rho(s)$ at different distances from the avalanche onset, ensemble averaged over an angular interval of $\Delta\theta=0.5^{\circ}$. The arrow is in the direction of approaching the avalanche. From left to right, $(\theta-\theta_{m})$ - values: $-14.4^{\circ}$ (green plusses), $-4.6^{\circ}$ (blue triangles), $-0.75^{\circ}$ (red squares), $-0.25^{\circ}$ (black dots); and during the avalanche (purple diamonds). Lines are fits to Eq. \ref{ps}, and the dashed line indicates a slope of $-2.5$.
Insets: (a) $\xi$, (b) maximum cluster size $S_{max}$ as a function of $\theta-\theta_{m}$, the distance to the avalanche onset. }
\label{fig:clusterdis}
\end{figure}

As a matter of fact, this complex dynamics leads to an interesting critical feature for the the cluster size distribution. As shown in Figure~\ref{fig:clusterdis} it develops larger and larger tails when approaching the avalanche onset. These distributions are indeed well fitted by a power law with an exponential cut off, the characteristic lengthscale of which increases sharply when approaching the avalanche onset:
\begin{equation} P(s) = \frac{1}{s^{\delta}} e^{-s/\xi},
\label{ps}
\end{equation}
where $s$ is the size of the cluster in particles. We have estimated $\delta = 2.5 \pm 0.25$ from the distribution just above onset (see dashed line) and the inset (a) displays the cluster scale $\xi$ as a function of the distance to avalanche onset. We observe a sharp upturn close to the avalanche, especially in the last 10 frames. Similarly, the ensemble-averaged size of the largest cluster also grows and  presents an upturn just before the onset of the avalanche (see inset (b) of Figure~\ref{fig:clusterdis}). 
Hence, despite their intrinsic marginal stability, and provided that the clusters are indeed quasi unidirectional, which was also established in~\cite{Lydie_PRE05}, the maximal size recorded here corresponds to clusters reaching system-spanning size at the onset of the avalanche.

\begin{figure}[t!]
 \centering
\includegraphics[width=0.85\columnwidth, trim = 0mm 0mm 5mm 5mm, clip]{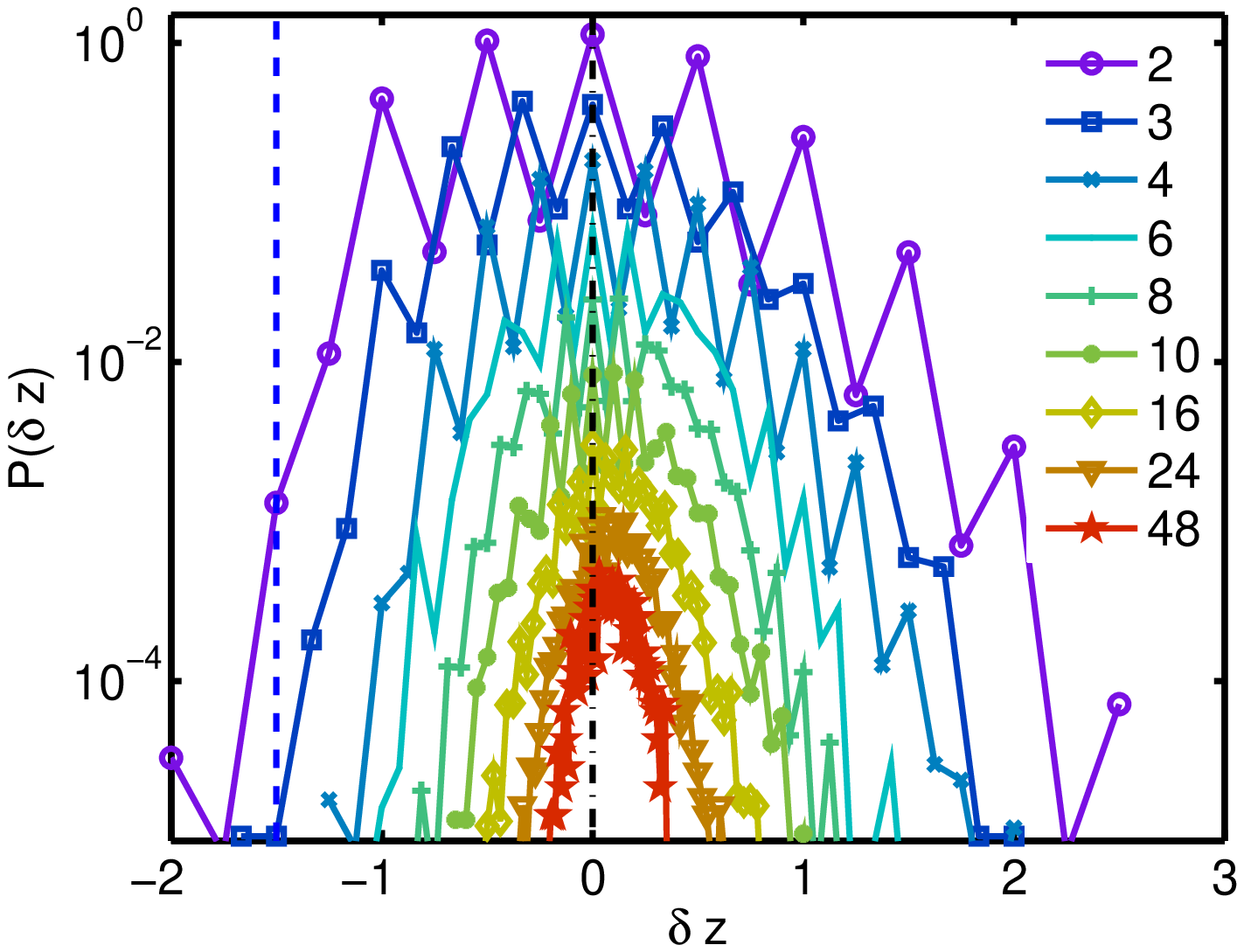}
\includegraphics[width=0.85\columnwidth, trim = 0mm 0mm 5mm 5mm, clip]{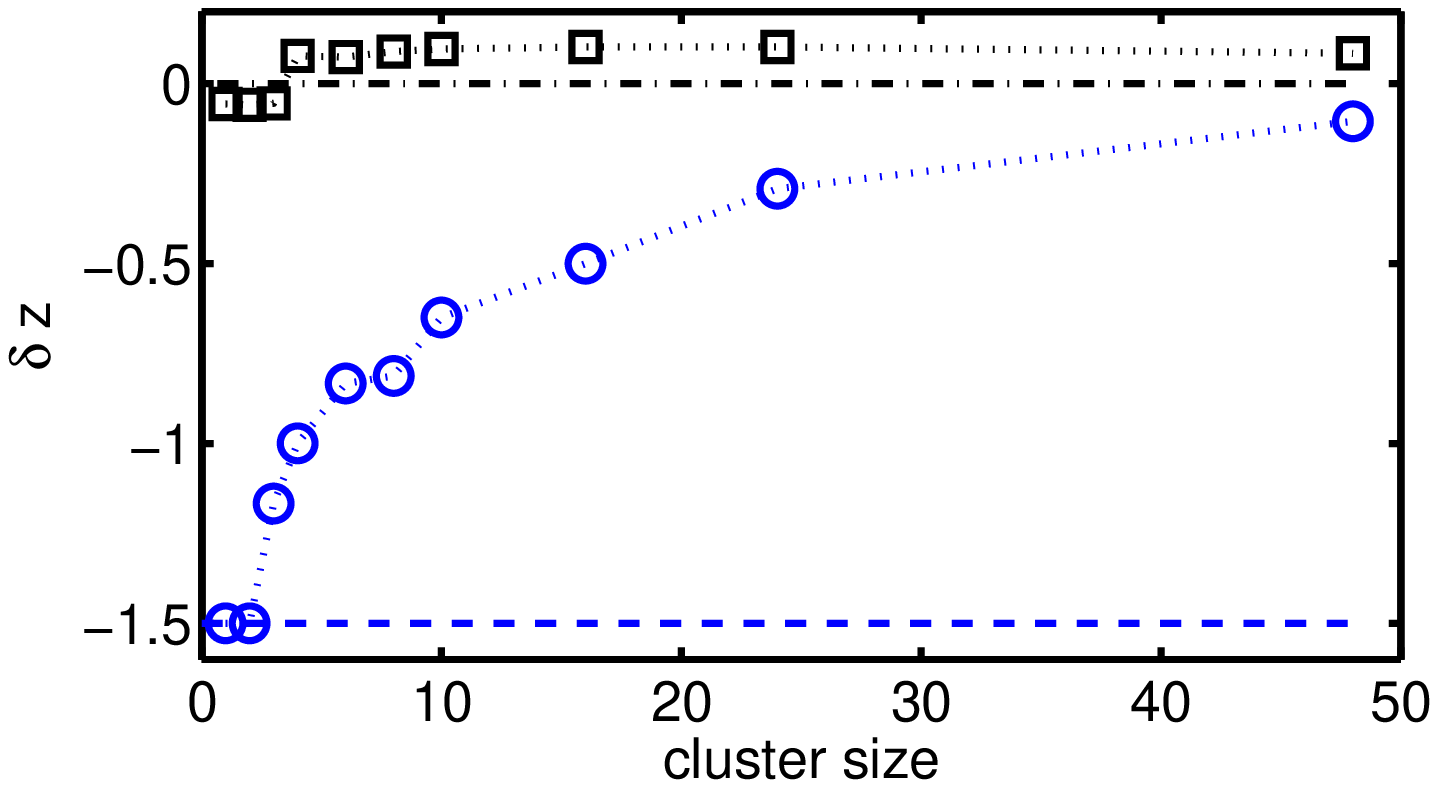}
\caption{Cluster size effects on stability:
\emph{a} -  Distribution of $\delta z$ for different cluster sizes, with binning that reflects the fundamental discreteness of the distributions. The curves have been shifted for clarity.
\emph{b} - In black (squares) is $\langle \delta z \rangle$ for different cluster sizes and the global stability criterion (dash-dot). The blue dots are the minimal $\delta z$ obtained for each cluster size; the dashed line is the marginal stability criterion for a single grain.}
\label{fig:clusters_stab}
\end{figure}

In the above discussion, we have assumed that the average contact number inside the clusters is roughly equal to the one for the whole pile. However given the increase of the exponential cut-off when approaching the avalanche, one suspects that the average contact number and to a lesser extent the average number of critical contacts actually depend on the cluster size. To elucidate this last point, we compute the distributions of $\delta z=z-2 n_m/d$ within clusters of a given size $N$ (see figure~\ref{fig:clusters_stab}a). Apart from discretization effects at small size and the shrinking of the width with $\sim N^{-1/2}$, the distributions progressively shift from negative $\delta z$ to a slightly positive one. Note also that the minimal value of $\delta z$ goes from the marginal local criterion for the smallest one (all studied clusters must be composed of locally stable particles) to close to the global stability criterion for the largest one (figure~\ref{fig:geniso_diag}b). Altogether these clusters appear to be dynamically selected according to their stability, which from a local mechanical proprety progressively builds up self consistently with their size towards the generalized isostaticity criterion.

The picture that emerges from this analysis is the following. The idea that the whole system is isostatic at the avalanche onset is too simple because it ignores the anisotropy and inhomogeneity of the pile, which transposes to the critical contacts. Such important effects of the anisotropy have been strongly emphasized before~\cite{Stephanie_PRE05,Lydie_PRE05} and it is clearly a too strong assumption to elude them~\cite{matthieu}. This suggests an even deeper history dependence of frictional piles, which calls for a refined description of the texture beyond the introduction of the number of critical contacts. For the isotropically compressed packings of \cite{Kostya,DOSslip}, it was also noted that only extremely slowly equilibrated packings unjam at $\delta z^{gen}=0$; for those packings, the critical contacts are indeed randomly distributed~\cite{our_obs}. The appearance of system-spanning marginally stable clusters is an intriguing mechanism of unjamming; one which cannot exist in frictionless systems if the packing structure and hence the local contact numbers are to remain homogeneous. This suggests that the frictional and the frictionless jamming transition may be more different than expected. Further insight in that matter could be gained from new experiments with two dimensional packings of photoelastic discs; in particular, the generalized isostaticity diagram could have direct relevance to the cyclic shear experiments of~\cite{bob_cyclic}. 

\section{Acknowledgements}
We would like to thank St\'ephanie Deboeuf for having shared her data with us. S. Henkes and C. Brito thank the FOM foundation for funding. O. Dauchot would like to thank the KNAW for the scientific opportunities given by his visiting professorship in Leiden. We acknowledge helpful discussions with Martin van Hecke.

\bibliography{avalanche_references}

\end{document}